# Nonlinear Strain-Mediated Magnetoelectric Coupling in Sub-Microscale Ni/BPZT Thin-Film Devices

Fanfan Meng[1]*, Emma Van Meirvenne[2], Federica Luciano[2], Xiangyu Wu[1], Yves Deblock[3], Kaustuv Banerjee[1], Peter Rickhaus[4], Florin Ciubotaru[1], Christoph Adelmann[1]*

[1]Imec, Leuven, Belgium
[2]Katholieke Universiteit Leuven, Leuven, Belgium
[3]Univ. Lille, CNRS, Centrale Lille, Univ. Polytechnique Hauts-de-France, UMR 8520 - IEMN – Institut d'Electronique de Microélectronique et de Nanotechnologie, F-59000 Lille, France
[4]Qnami AG, Muttenz, Switzerland

## Abstract

Strain-mediated magnetoelectric (ME) heterostructures enable electric-field control of magnetism and are promising for ultra-low-power spintronic logic. Yet achieving spatially selective, low-voltage control in thin films and quantifying ME coupling across the full ferroelastic loop remains challenging. Here, we investigate sub-micrometer Ni/BPZT thin-film devices with laterally patterned gates that localize in-plane strain beneath the Ni stripe and modulate its magnetization. We use anisotropic magnetoresistance to measure magnetization changes across the ferroelastic loop under different magnetic bias fields. Combined with Multiphysics strain simulations and micromagnetic modeling, this provides a quantitative framework that captures the convolution of ferroelastic and magnetoelastic nonlinearities and provides critical insight for device design, while enabling multi-state, bias-field-free magnetization control for non-conventional computing. The extracted coupling coefficient in linear range is 1.3 mT V$^{-1}$ across a 700 nm gap, with a clear pathway to improving voltage efficiency through device scaling, establishing a scalable CMOS-compatible platform for energy-efficient spintronic devices.

## Introduction

Spintronic devices encode information in the direction of magnetization and are regarded as leading candidates for beyond-CMOS technologies because of their potential for ultra-energy-efficient computing and compatibility with non-traditional computing paradigms.[1] Although magnetic random-access memory (MRAM) has reached commercial availability, spintronic devices have yet to be adopted in computing systems. A key limitation lies in the established mechanisms that convert charge information to spin, namely spin orbit torque or spin transfer torque, which require current densities in the range of $10^5$ to $10^7$ A cm$^{-2}$ and therefore incur substantial energy consumption.[2]

To improve the energy efficiency of charge-spin conversion and to enable logic applications,[3,4] magnetoelectric (ME) multiferroic materials have attracted growing interest for their ability to control magnetism with electric fields at significantly lower energy.[5] Multiferroics possess more than one ferroic order, such as ferroelectricity (FE) and (anti)ferromagnetism (FM), that are coupled.[6] Electric-field control of magnetism in multiferroics can be realized using single-phase materials [7–9], though these often exhibit weak FM ordering and Curie temperatures below room temperature. Alternatively, the ME coupling can be achieved in multiferroic heterostructures through four primary mechanisms: ionic migration,[10] charge accumulation,[11,12] exchange bias,[13,14] and strain transfer.[15–17] It is worth noting that exchange-driven coupling has achieved low switching voltage of 200 mV with 10 nm thick Bi$_{0.85}$La$_{0.15}$FeO$_3$, but further down-scaling is precluded by leakage in ultra-thin films.[8] Consequently, strain-mediated systems, where a magnetostrictive ferromagnet is elastically coupled to a piezoelectric layer, have garnered attention for their exceptionally high coupling coefficients, robust room-temperature operation, wide materials choice, and compatibility with device miniaturization.[17]

Most strain-driven ME devices reported to date employ continuous[15,18] or patterned[19,20] magnetostrictive thin films on bulk single-crystal piezoelectric substrates contacted by top and bottom electrodes. While they demonstrate strong coupling[15] and switching of nanomagnets[19], these approaches require operating voltages on the order of hundreds of volts, usually impose uniform strain on all elements without control over individual magnets, and are difficult to integrate with standard semiconductor fabrication processes, making them less relevant for device-level studies.

For strain-driven ME devices to become truly relevant for microelectronic applications such as logic or memory, several key requirements must be addressed. First, piezoelectric thin films need to be deposited on silicon substrates to enable scalable fabrication. Second, the magnetic thin film must be patterned into nanoscale elements that can be individually controlled. Third, quantitative determination of the magnetization response and the corresponding ME coupling across the full ferroelastic range, not just the linear piezoelectric regime, is essential to establish device specifications and benchmarks. The quantification performed on extended, unpatterned thin films is not sufficient, as the strain profile differs fundamentally from that in scaled nanomagnets. Finally, a framework is required to interpret this quantification, including the efficiency of strain transfer to the nanomagnet and the nonlinear contributions from both the ferroelastic strain and magnetoelastic coupling. In particular, quantitative mapping of magnetization rotation throughout the full ferroelastic loop is critical for device optimization. However, only a few studies [16,21,22] have reported using thin film piezoelectrics to alter the magnetization of nanoscale magnetic elements. When addressed, these works typically focus only on the initial and final magnetization states under applied voltage and relied mainly on qualitative microscopy studies. Others have been restricted to the linear piezoelectric regime, without resolving the nonlinear contributions from both ferroelastic strain and magnetoelastic coupling.

In this work, we aim to bridge these gaps by investigating strain-mediated ME coupling in a sub-micron Ni stripe patterned on a Ba-substituted Lead Zirconate Titanate (PZT) thin film. Using laterally patterned electrodes on the top surface of the piezoelectric layer, we generate localized, in-plane strain directly beneath the magnetic element, enabling spatially selective, low-voltage control of magnetization. We then track the magnetization response across the entire ferroelastic switching loop under a range of transverse magnetic bias fields via anisotropic magnetoresistance measurements (AMR), AMR provides well-established, quantitative access to magnetization rotation in patterned ferromagnets, offering information that cannot be obtained from microscopy alone. This enables full quantification of the magnetization response and ME coupling. To elucidate the coupling mechanism, particularly the convolution of nonlinearities arising from both ferroelastic strain and magnetoelastic coupling, we combine COMSOL Multiphysics® strain simulations, micromagnetic modeling, and scanning nitrogen–vacancy (NV) magnetometry. NV magnetometry maps directly visualize domain evolution under voltage bias, serving as an independent probe that validates both the AMR and simulation analysis.

# Results and Discussion

## 1. Device and measurement configurations

A schematic of the device is shown in **Figure 1a**. A 550 nm-thick BPZT [$(Ba_{0.1}Pb_{0.9})(Zr_{0.52}Ti_{0.48})O_3$] film was deposited by pulsed laser deposition on a SiO$_2$/Si substrate, with a 10 nm LaNiO$_3$ (LNO) seed layer and was chemical mechanically polished to 370 nm. A $16\ \mu m \times 1\ \mu m \times 30\ nm$ Ni stripe was patterned on the PZT using electron-beam lithography, sputtering, and lift-off. We subsequently patterned a pair of Au gate electrodes positioned 700 nm away from each edge of the stripe to apply lateral electric fields, along with four Au contact pads on the stripe for four-probe anisotropic-magnetoresistance (AMR) measurements. The completed device is shown in the top-view scanning-electron-microscopy (SEM) image in **Figure 1b**; further fabrication details are provided in Methods.

**Figures 1c and 1d** present finite-element simulations using COMSOL Multiphysics® of the electric field distribution when 30 V is applied to both gates while the Ni stripe is grounded. Unlike prior works where a bottom electrode is grounded and the electric field is dominated by out-of-plane component, our device utilizes fringing fields that extend from the Au gates toward the stripe; the in-plane component $E_y$, dominates and is nearly uniform across the gap, while the out-of-plane component $E_z$, is confined to the electrode edges and is roughly half as large. Through the direct piezoelectric effect, these fields deform the PZT layer. **Figures 1e and 1f** show that the PZT expands laterally between the gates, imposing a localized, relatively uniform compressive strain along $y$-direction to the Ni stripe. Poisson contraction induces an out-of-plane expansion approximately one-third as large. This electrostatically induced strain underpins the magnetoelectric coupling in the device.

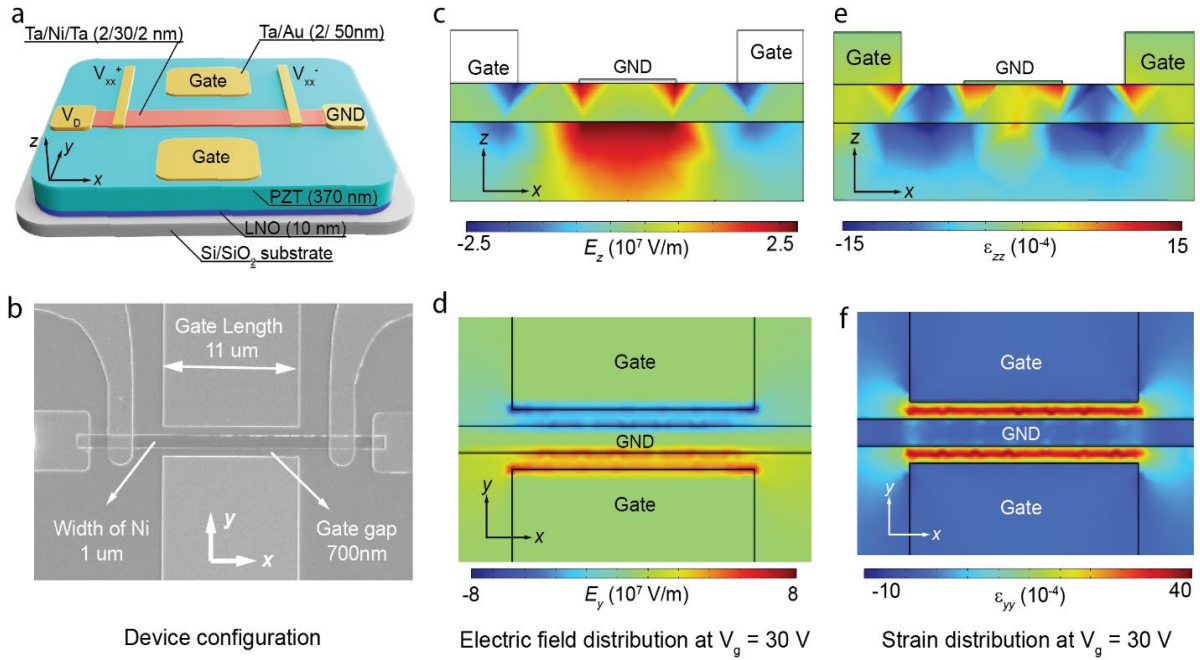

*Figure 1. Device structure and operation. (a) Schematic of the sub-microscale Ni/BPZT device with lateral gate pairs and four contacts for magnetoresistance measurements. The Ta/Ni/Ta (2/30/2) nm stripe is patterned on top of a 370 nm PZT film. (b) Top-view SEM image of the fabricated structure, showing a Ni stripe with 1 μm width and 16 μm length, and a 700 nm-gap between gate electrodes and Ni stripe. (c, d) Simulated electric field distribution at 30 V: (c) z-component ($E_z$) in the xz cross-section showing fringing fields penetrating the PZT layer, and (d) y-component ($E_y$) in the xy plane highlighting lateral field confinement between gates. (e, f) Simulated strain distribution at 30 V: (e) z-component ($\varepsilon_{zz}$) in the xz plane indicating vertical strain localization beneath the gate edges, and (f) y-component ($\varepsilon_{yy}$) in the xy plane showing lateral strain concentration along the Ni stripe between the gates.*

## 2. Magnetic characterization of the Ni stripe

We first characterize the magnetoresistance of the Ni stripe under applied magnetic field, in the absence of gate voltage, to establish a baseline. **Figure 2a** shows the anisotropic magnetoresistance (AMR) hysteresis loops measured under magnetic fields applied along the three principal axes. AMR arises from different degrees of scattering of spin–orbit coupled carriers, depending on the relative orientation between the magnetization and the current direction[23,24]. Phenomenologically, the resistance as a function of magnetization angle $\theta$ relative to the current direction is given by
$$R(\theta) = R_\perp + (R_\parallel - R_\perp)\cos^2(\theta),$$
where $R_\perp$ and $R_\parallel$ are the resistances for magnetization perpendicular and parallel to the current, respectively. From the saturation values in hysteresis, we obtain $R_\perp = 138.4\ \Omega$, and an AMR ratio defined as $(R_\parallel - R_\perp)/R_\perp = 1.29\%$.

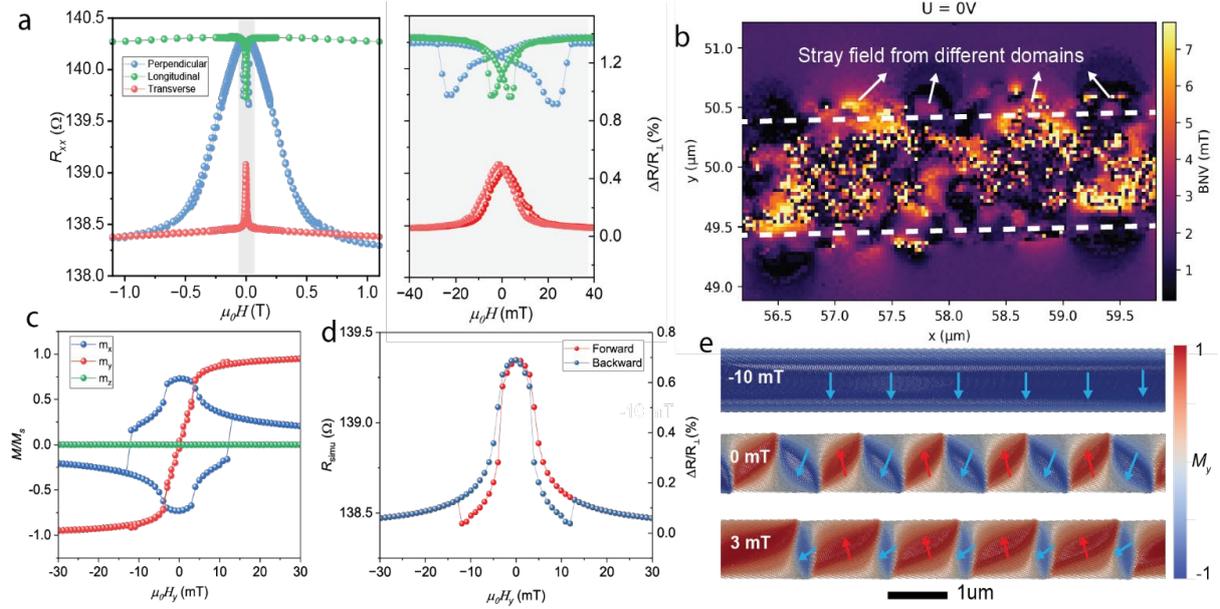

Figure 2. (a) Anisotropic magnetoresistance (AMR) measured under magnetic fields applied along the three principal directions, showing full hysteresis loops from -1.1 T to 1.1 T and backwards. (b) Scanning nitrogen-vacancy magnetometry (SNVM) map of the stray magnetic field above the Ni stripe at remanence, indicating a multi-domain state. White dashed lines indicate the location of the Ni stripe. (c) Micromagnetic simulation of the magnetization components ($m_x$, $m_y$, $m_z$) as a function of transverse magnetic field. (d) Simulated AMR loop under transverse magnetic field, reproducing the observed hysteresis characteristic. (e) Simulated magnetic domain configurations (color-mapped by $M_y$) under different transverse bias fields (−10 mT, 0 mT, and 3 mT), illustrating field-induced domain evolution.

Notably, in the transverse hysteresis, the resistance only reaches approximately 50% of the full AMR range (**Figures 2a, red loops**), indicating that the remanent magnetization is not fully aligned with the current direction, which coincides with the long axis of the Ni stripe. This suggests the presence of transverse magnetic anisotropy that counteracts the shape anisotropy of the elongated stripe. Consistent with this, the scanning nitrogen–vacancy magnetometry (SNVM) map, which characterizes the magnetic stray fields[25] of the Ni stripe, in **Figure 2b**, clearly shows multidomain states at remanence, confirming the presence of transverse anisotropy in the Ni stripe.

To corroborate these observations, we carried out micromagnetic simulations using MuMax3 (**Figures 2c–e**), introducing a uniaxial transverse anisotropy of $K_u = 2000\ J/m^3$. (see Methods for additional parameters used). The simulations reproduce the ground-state domain pattern and the AMR hysteresis, yielding a maximum simulated resistance change of 0.7 %, close to the 0.6 % measured experimentally. As the magnetic field decreases from −1 T to about −5 mT, the magnetization rotates coherently; below −5 mT the stripe breaks into alternating domains. At remanence, the central magnetization vector is tilted by roughly 65° from the stripe axis. Further reversing the field causes domains whose magnetization opposes the field to shrink and reorient toward the new field direction.

The required transverse anisotropy may stem from built-in stress[26] or from oblique-angle deposition[27] during Ni sputtering.

## Voltage-driven magnetization control

Building on the baseline magnetic characterization, we next investigate the strain-mediated magnetoelectric (ME) coupling under transverse magnetic bias fields. In the setup shown in **Figure 3a**, a constant field $B_y$ presets the stripe magnetization, while a bi-directional gate sweep from ±30 V induces localized strain in the PZT. The resulting magnetization change is probed via anisotropic magnetoresistance (AMR) measurements. **Figure 3b** shows butterfly-shaped AMR hysteresis loops, resembling the ferroelastic strain signature typical of ferroelectric materials. Capacitance–voltage (C–V) data in **Figure 3c** confirm ferroelectric switching with coercive voltages at ±2 V. Consequently, the ferroelectric nature of the PZT must be included in the analysis, and linear piezoelectric models alone are insufficient to capture the observed magnetization dynamics.

To understand how ferroelastic strain influences magnetization, we model the strain evolution in COMSOL Multiphysics by incorporating a combination of the Jiles–Atherton (J–A) polarization model[28] and an electrostriction model[29], which captures the nonlinear, hysteretic response of PZT. While the model uses generic PZT parameters from the COMSOL library, the goal is not to reproduce exact experimental strain values, but to understand how such strain profiles impact magnetization in the Ni stripe. **Figure 4a** shows the simulated principal strain components ($\varepsilon_{xx}$, $\varepsilon_{yy}$, and $\varepsilon_{zz}$) within the Ni layer as a function of gate voltage. The transverse in-plane strain $\varepsilon_{yy}$ dominates, peaking at ±30 V and reversing sign between zero and the coercive voltages. The out-of-plane strain $\varepsilon_{zz}$ is opposite in sign but similar in magnitude to $\varepsilon_{yy}$, consistent with Poisson-driven deformation, while the longitudinal strain $\varepsilon_{xx}$ remains negligible.

The strain couples to magnetization through an effective magnetoelastic field[30]:

$$H_{mel} = -\frac{2}{\mu_0 M_s} \begin{pmatrix} B_1 \varepsilon_{xx} m_x + B_2(\varepsilon_{xy} m_y + \varepsilon_{zx} m_z) \\ B_1 \varepsilon_{yy} m_y + B_2(\varepsilon_{xy} m_x + \varepsilon_{yz} m_z) \\ B_1 \varepsilon_{zz} m_z + B_2(\varepsilon_{zx} m_x + \varepsilon_{yz} m_y) \end{pmatrix}.$$

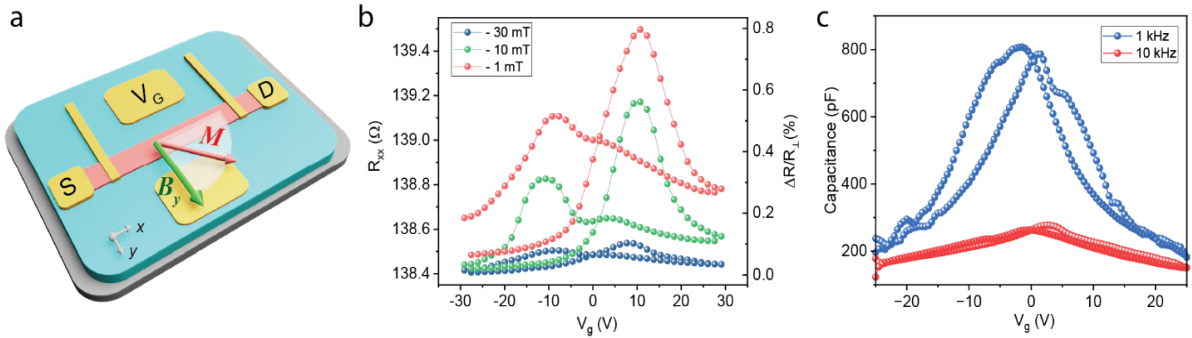

Figure 3. (a) Schematic of the measurement configuration. A transverse magnetic bias field $B_y$ is applied to preset the magnetization direction of the Ni stripe. A bi-directional voltage sweep (−30 V to +30 V) is then applied to the lateral gate electrodes to modulate strain in the underlying PZT. The magnetization response of the Ni stripe is monitored via the anisotropic magnetoresistance (AMR) effect. (b) Anisotropic magnetoresistance (AMR) measured during the voltage sweep for three different magnetic bias fields (−30 mT, −10 mT, and −1 mT). (c) Capacitance–voltage (C–V) measurements performed at 1 kHz and 10 kHz confirm ferroelectric switching behavior.

with $B_1$ and $B_2$ the first and second magnetoelastic coupling constants, and $\varepsilon_{ij}$ the components of the mechanical strain tensor. We estimate the magnetoelastic coupling constant $B_1$ by analyzing the AMR response at a bias field of –30 mT (blue curve in **Figure 3b**). In this case, the magnetization is expected to undergo coherent rotation without domain formation. We convert the measured magnetoresistance to the average magnetization angle using Equation (1). **Figure 4b** plots the average angle versus gate voltage. As the gate voltage increases from -30 V to +10 V, the magnetization rotates from −86° to −72°, corresponding to a reduction in effective transverse field reduction of 40 mT. Because the magnetization is nearly saturated along the y-direction, the contribution from $m_x$ and $m_z$ are negligible, and all shear strain components are negligible compared to the normal strain, simplifying the ME field to

$$H_{mel} \approx -\frac{2}{\mu_0 M_s}\begin{pmatrix} 0 \\ B_1 \varepsilon_{yy} m_y \\ 0 \end{pmatrix}.$$

From this relation we determine an effective magnetoelastic constant of $B_1 = 7.1 \times 10^6$ J/m³, which is used in all subsequent micromagnetic simulations.

The three normal strain components ($\varepsilon_{xx}$, $\varepsilon_{yy}$, and $\varepsilon_{zz}$) obtained from COMSOL (Figure 4a) were imported as spatially uniform inputs into micromagnetic simulations to model the magnetization evolution under a series of applied strains (see Methods). Figures 3b and 4c compare measured and simulated AMR loops under transverse magnetic bias fields of −30 mT, −10 mT, and −1 mT, while **Figure 4d** shows the corresponding y-component of the magnetoelastic field, $H_{mel,y}$. At −30 mT, the simulation reproduces the butterfly shape and maximum resistance change (~0.1%) observed experimentally. In the quasi-linear region (from 4 V to 30 V), $H_{mel,y}$ varies by ≈ 37 mT, giving an effective magnetoelectric coefficient of 1.3 mT/V.

At an intermediate bias field of −10 mT, both simulation and experiment capture a broader butterfly-shaped AMR loop. The simulated maximum resistance change (0.8%) slightly exceeds the experimental value (0.6%), likely due to the idealized strain profile or uncertainties in the material parameters, yet the overall trend remains consistent. For large negative gate voltages (−30 V to −5 V), the measured magnetoresistance almost coincides with the −30 mT trace, and the simulated $H_{mel,y}$ is only marginally lower, indicating that the strain is sufficient to keep the magnetisation quasi-saturated even at the reduced bias. The main discrepancy arises at low voltages (< 10 V), where diminishing strain weakens $H_{mel,y}$. Under −10 mT, this weaker field can no longer counteract the shape anisotropy, so the magnetization rotates more easily toward the stripe axis. This leads to a steeper rise in resistance and further reduction of $H_{mel,y}$, making it more difficult to rotate the magnetization back towards the transverse direction, thus broadening the AMR loop.

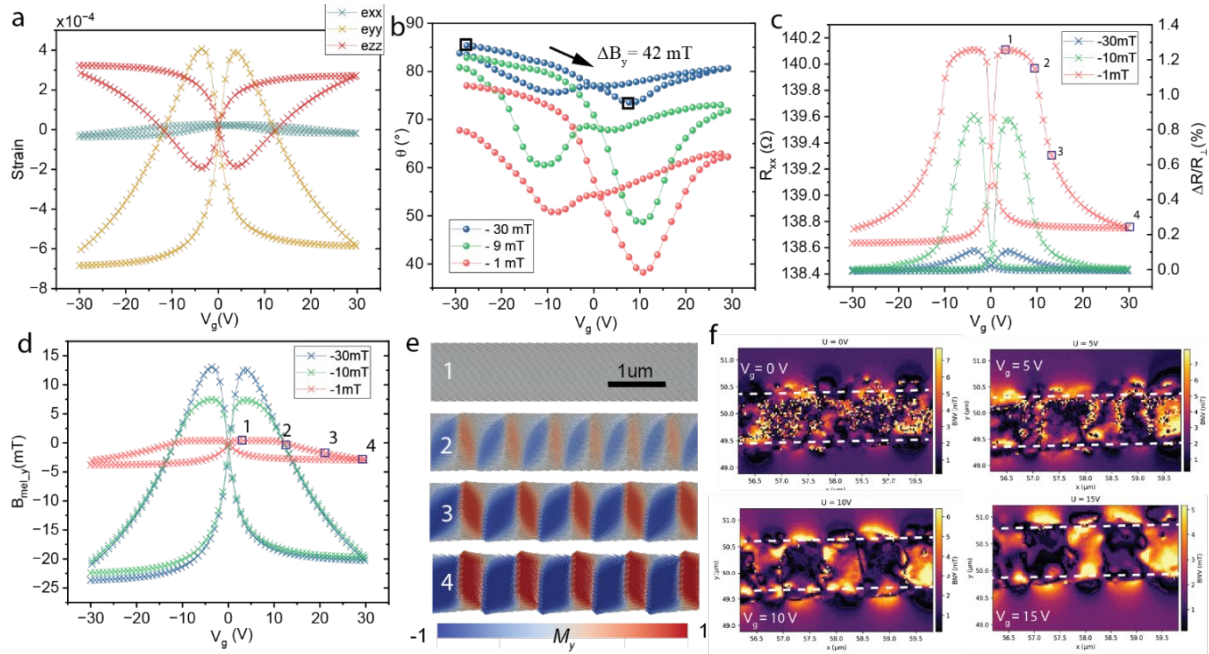

*Figure 4. (a) Simulated strain components ($\varepsilon_{xx}$, $\varepsilon_{yy}$, and $\varepsilon_{zz}$) in the Ni stripe as a function of applied gate voltage $V_g$, using a Jiles–Atherton ferroelectric polarization model in COMSOL Multiphysics. (b) Average magnetization angle θ extracted from experimental AMR data under different transverse magnetic bias fields. (c) Simulated AMR loops under transverse bias fields of −30 mT, −10 mT, and −1 mT, using full micromagnetic simulations with strain input from COMSOL Multiphysics. (d) Simulated magnetoelastic effective field component $B_{mel,y}$, calculated from strain and magnetization configurations. (e) Snapshots of the simulated magnetization $M_y$ profile along the Ni stripe under a bias field of −1 mT, showing transition from uniform to alternating domain states as gate voltage increases. (f) NV magnetometry scans of the Ni stripe under four gate voltages (0 V, 5 V, 10 V, and 15 V), measured without applying magnetic bias field. The stray field patterns evolve with voltage, consistent with domain nucleation and expansion.*

At −1 mT bias, both experiment and simulation show the largest resistance change, although the simulation slightly overestimates the magnitude. This pronounced response does not arise from a stronger magnetoelastic field; in fact, $H_{mel,y}$ is only about one-eighth of the value obtained at −30 mT and −10 mT (Figure 4d). In micromagnetic simulations, in "state 1" (Figure 4e), under small gate voltage, the magnetization relaxes into a longitudinal configuration along the Ni stripe, resulting in high resistance. As voltage increases, even this modest ME field is sufficient to trigger a transition into alternating domain states with ±y magnetization components (state 2). Further increases in $H_{mel,y}$ aligns these domains more firmly along ± y directions (state 3 and 4), generating a larger stray field captured in the NV magnetometry scan (Figure 4f) and leading a sharp resistance drop. The relatively low but persistent ME field (Figure 4d, red curve) reflects that the stripe's soft magnetic configuration and transverse anisotropy allow significant magnetization reorientation without requiring large ME fields. These results underscore the importance of engineering initial magnetic states for efficient voltage-driven control, providing valuable insights for low-power spintronic device design.

## Conclusion

In this work, we demonstrate localized ferroelastic strain-driven, magnetoelectric (ME) coupling in a sub-micrometer Ni/PZT thin film device using laterally patterned electrodes. This design enables individual magnet control without the high-voltage (~100s V) typically required for bulk single-crystal piezoelectric substrates. By integrating AMR measurements with COMSOL Multiphysics strain modelling and MuMax3 micromagnetic simulations, we quantitatively determined an effective magnetoelectric coupling coefficient of approximately 1.3 mT/V in the linear range and extracted a magnetoelastic coupling constant $B_1 = 7.1 \times 10^6$ J/m³. While this voltage-based coefficient is lower than ~10 mT/V reported for 20 nm BiFeO$_3$ thin films at 500 mV switching[8], our electric-field-based ME coupling is approximately five times higher, and with a current gate gap of 700 nm (compared to 20 nm in BFO), there is substantial opportunity to enhance coupling at low operating voltages by down-scaling to the sub-100 nm regime. AMR provides quantitative access to magnetization rotation across the full ferroelastic loop, and independent nitrogen–vacancy magnetometry maps confirm the domain evolution inferred from transport and simulations. Together, these approaches reveal that strain-induced domain evolution yields the largest MR response at a 1 mT bias field, highlighting the decisive role of the engineered remanent state and providing a pathway toward bias-field–free magnetization rotation. The pronounced nonlinearity introduced by both ferroelastic strain and ME coupling highlights the promise of ME structures for non-Boolean logic implementations, where multi-state behavior can be exploited.

A promising route to further improve voltage coupling efficiency is to develop vertical ME structures, by fully etching both the piezoelectric and magnetic layers. This would reduce substrate clamping of the piezoelectric film and allow top-down gate electrodes to exploit the larger out-of-plane $d_{33}$ piezoelectric coefficient instead of the in-plane $d_{31}$. Prior work has demonstrated that employing PMN-PT membranes with vertical gating to reduce clamping holds strong promise for enhancing ME coupling, although a scalable approach remains to be established.[31] In particular, integrating perpendicular magnetic anisotropy materials in such vertical structures would further leverage the out-of-plane $d_{33}$ response to achieve efficient, perpendicular magnetization control rendering higher storage or logic density. At the same time, employing materials with enhanced magnetostrictive properties[17,32–34] or utilizing advanced relaxor ferroelectrics with large piezoelectric responses[35–37] could significantly improve the overall device performance and reduce the required operating voltage.

## Methods:

550 nm thick BPZT [(Ba0.1Pb0.9)(Zr0.52Ti0.48)O3] was deposited by pulsed laser deposition (PLD) using a Solmates SIP-700 system at 550 °C in an O2 atmosphere. A 10 nm thick LaNiO3 (LNO) layer also deposited by PLD was used as a buffer layer for the piezoelectric BPZT. The full stack is the following: Si (100) substrate / SiO2 (400 nm) / Ti (20 nm) / Pt (70 nm) / LNO (10 nm) / BPZT (550 nm). The Ti/Pt bottom electrode was deposited by e-beam evaporation. Post deposition, a chemical mechanical polishing (CMP) was performed on the BPZT thin film to reduce the roughness. The samples were polished in a Mecapol E460 system, using an Optivision 4548 soft pad and an ILD3013 (pH 10-11) SiO2 abrasive slurry. The CMP conditions for BPZT polishing have been optimized in a previous study.[38] The table/head speeds and downforce were set at 50 rpm / 50 rpm and 1.5 psi (~0.1 mbar), respectively. The final root-mean-square (RMS) roughness obtained was below 1 nm and the final BPZT thickness was 370 nm. To remove the slurries residues, a post-CMP standard cleaning was performed, it included 10 minutes in SC-1 solution (NH4OH:H2O2:H2O = 1:1:5) at 75 °C and 5 minutes in isopropanol in an ultrasonic bath.

Scanning NV magnetometry (SNVM) measurements were performed under ambient conditions using Qnami ProteusQ system. The NV flying distance is estimated to be 50 nm. SNVM enables direct mapping of local magnetic stray fields via optical detection of the Zeeman shift in the NV spin levels, with a spatial resolution corresponding to the NV-to-magnet distance. During measurement, gate voltages were applied to the device via a customized PCB integrated into the Qnami system. This setup allows nanoscale imaging of voltage-tunable magnetic textures relevant to spintronic and magnetoelectric devices. Pixel size of 30x30nm$^2$ and pixel integration time =0.42s.

Anisotropic magnetoresistance measurements were performed using a CRX-VF cryogen-free, closed-cycle probe station equipped with a 2.5 T vertical superconducting magnet. An in-plane magnetic field was applied using a custom-designed sample holder, and all measurements were conducted at a temperature of 295 K. A standard four-terminal AC lock-in technique was employed, with a constant input current amplitude of 10 µA at a frequency of 111 Hz. The gate voltage was applied using a Keithley 2620 source meter.

The strain components were calculated as a function of applied electric potential using steady-state finite element simulations in COMSOL Multiphysics®. The mechanical material parameters for PZT were taken from the COMSOL library, and the ferroelectric parameters were measured internally[38]. Except for the lateral size of the PZT substrate, the model geometry matched the actual fabricated device. A nickel strip was placed on top of 16.5 µm × 50 µm × 370 nm PZT layer. The PZT sides were clamped to reflect the much larger size of the real substrate, while a soft bottom layer was used to reduce substrate clamping and improve strain transfer. A voltage sweeps from –30 V to +30 V was applied in 0.3 V increments to capture ferroelectric switching behavior. The Electrostatics and Electrostriction modules were used to calculate the electric field and polarization in the PZT, with switching behavior captured via the Jiles–Atherton hysteresis model, which calculates the hysteretic polarization. The saturation polarization and polarization reversibility are set to 0.4 C/m² and 0.2 respectively. The pinning loss is 8.06 MV/m, interdomain coupling is 4.2 Mm/F and the domain wall density is set to 0.641 MV/m. The electrostrictive strain tensor was then coupled to the Solid Mechanics module to simulate mechanical deformation, and strain transfer from the PZT layer to the nickel strip was computed. The average in-plane and out-of-plane strain components (e.g., $\varepsilon_{xx}$, $\varepsilon_{yy}$, $\varepsilon_{zz}$) in the nickel layer were extracted for each voltage step and subsequently used as input for micromagnetic simulations.

Micromagnetic simulations were performed using MuMax3[39]. The Ni layer was modeled with an exchange stiffness of $A_{\text{ex}} = 6 \times 10^{-12}$ J/m, a saturation magnetization of $M_s = 400 \times 10^3$ A/m, and a Gilbert damping constant $\alpha = 0.02$. The exchange length, defined as $l_{\text{ex}} = \sqrt{2A_{\text{ex}}/(\mu_0 M_s^2)}$, was calculated to be 7.73 nm. The simulation cell size was set to 7 nm, ensuring sufficient spatial resolution. The average strain components in the Ni layer, computed from COMSOL simulations, were imported as a series of uniform values using the built-in magnetoelastic modules exx (), eyy(), and ezz () APIs in MuMax3. The resultant magnetization configuration under each strain condition was obtained using the relax () and minimize () functions.

## Acknowledgements:

This work was supported by IMEC's Industrial Affiliation Program on Exploratory Logic Devices F. Meng was supported by European Union (EU)'s Horizon Europe Framework Programme under the Marie Skłowdowska-Curie Postdoctoral Fellowship Action 2023 'All Magneto-Electric Spin Logic Gates' — ALLME with grant agreement No 101106745. E. Meirvenne and F. Luciano were supported by the Research Foundation Flanders (FWO) through grants No 1SH4Q24N and 1183722N, respectively. This work has been partially undertaken with the support of IEMN CMNF electron beam facilities and supported by the French Renatech network. We gratefully thank Cesar Javier Lockhart de la Rosa and Gouri Sankar Kar for their support in coordinating the imec resources.